\documentclass[prd,aps,showpacs,groupedaddress,eqsecnum,notitlepage,nofootinbib,superscriptaddress]{revtex4-2}
\usepackage{subcaption}
\usepackage{graphicx}

\usepackage{tikz,siunitx,mwe}
\usepackage{tensor}
\usepackage{epstopdf}
\usepackage{amsmath}
\usepackage{amsfonts}
\usepackage{slashed}
\usepackage{amssymb}
\usepackage{enumerate, color}
\usepackage{subcaption}
\usepackage{lipsum}
\usepackage{color}
\usepackage{graphicx,bm}
\usepackage[utf8]{inputenc}
\usepackage{eurosym}
\usepackage{scalerel}
\usepackage{float}
\usepackage{accents}
\usepackage[colorlinks=true,pdfstartview=FitV,linkcolor=blue,citecolor=blue,urlcolor=blue,breaklinks=true]{hyperref}
\usepackage{orcidlink}

\begin{document}

\title{Geometric and Topological Aspects of Quadrupoles of Disclinations: Conformal Metrics and Self-Forces}


\author{A. M. de M. Carvalho\,\orcidlink{0009-0006-3540-0364}}
\email{alexandre@fis.ufal.br}
\affiliation{Instituto de F\'{\i}sica, Universidade Federal de Alagoas, 57072-970, Macei\'o,  AL, Brazil.}
\author{G. Q. Garcia\,\orcidlink{0000-0003-3562-0317}}
\email{gqgarcia99@gmail.com}
\affiliation{Centro de Ci\^encias, Tecnologia e Sa\'ude, Universidade Estadual da Para\'iba, 58233-000, Araruna, PB, Brazil.}
\author{C. Furtado\,\orcidlink{0000-0002-3455-4285}}
\email{furtado@fisica.ufpb.br}
\affiliation{Departamento de F\'isica, Universidade Federal da Para\'iba, 58051-970, Jo\~ao Pessoa, PB, Brazil. }
\begin{abstract}
We study the geometric and physical effects of quadrupolar configurations of disclinations using a conformal metric approach in $(2+1)$ dimensions. Two cases are considered: a linear quadrupole, inducing anisotropic curvature with a $\cos(2\theta)$ as profile, and a square quadrupole, yielding a more isotropic field with higher angular harmonics. We solve the Poisson equation for the conformal factor and compute the corresponding Green functions. Using these configurations, we evaluated the electrostatic and magnetostatic self-energies and self-forces for linear sources. The results reveal how symmetry and curvature influence self-interaction effects, with the magnetostatic self-force exhibiting a sign reversal compared to the electrostatic case. Connections with previous models of dislocations and cosmic strings are discussed, with potential applications in graphene, nematics, and gravitational analogs.
\end{abstract}

\maketitle

\section{Introduction}
Topological defects are central to a broad range of physical systems, from the early-universe cosmology to the condensed matter physics~\cite{DeserJackiw1984, DeserJackiw1989, katanaev1992theory}. These defects, such as cosmic strings, disclinations, and dislocations, can be interpreted as non-trivial geometries embedded in otherwise flat spacetimes, and their influence on classical and quantum fields has motivated deep theoretical developments~\cite{Fumeron2021, kleinert1989gauge, gopalakrishnan2006curvature}. In particular, topological defects induce long-range modifications in the electromagnetic, gravitational, and scalar fields, giving rise to self-interaction phenomena. Even when the local curvature vanishes, global geometric structures impose boundary conditions that generate nonlocal effects. These self-force and self-energy effects have been studied in conical geometries~\cite{lin, sou, mex}, global monopoles~\cite{eug1, mello2012}, and lower-dimensional gravity scenarios~\cite{furtado1997, DeserJackiw1984, DeserJackiw1989}. Furthermore, the deep connection between topology and electromagnetism has historical roots in classical studies on magnetic monopoles and global gauge formulations~\cite{dirac1931quantised, wu1975concept, eguchi1980gravitation}, and continues to be actively explored in contemporary contexts such as singular optical fields and electromagnetic vortices~\cite{dennis2009singular, berry2009optical, abulafia2023}.

The geometric theory of defects~\cite{katanaev1992theory, kleinert1989gauge} builds a powerful bridge between elasticity and gravitation by modeling disclinations and dislocations as sources of curvature and torsion in an effective manifold. Several condensed matter systems have been described by the geometric theory of defects. For instance, defects in the $^{3}He-A$ superfluid were treated as a gravitational analogue in refs.~\cite{volovik1998simulation, carvalho2025scaterring}, where the dysgiration defect was compared to the cosmic string. In ref.~\cite{candemir2023linear}, the authors studied the linear and nonlinear optical properties in the $GaAs$ quantum dot through of the geometric theory of defects. The presence of pentagon/heptagon rings in the graphene lattice has been a subject to the geometric theory of defects, where these disclinations are related to the curvature of graphene~\cite{vozmediano2010gauge}. In addition to this, the presence of dislocations (pentagon-heptagon pairs) is related to the torsion of the spacetime~\cite{mesaros2010parallel, mesaros2009berry}. As a consequence, recent developments extend these ideas to structured defect distributions, such as dipoles and quadrupoles, which are found in liquid crystals~\cite{Fumeron2021}, graphene lattices~\cite{Guinea2010} and nematic shells~\cite{gopalakrishnan2006curvature}. However, most investigations have focused on isolated defects, leaving the effects of complex defect structures, such as quadrupoles of disclinations, less explored.

Our focus is on the quadrupole configuration, which is the symmetric arrangement of four disclinations. From a mathematical perspective, the formulation of a quadrupole of disclinations involves solving the Poisson equation for the conformal factor, which encodes the curvature effects induced by the defect structure~\cite{katanaev1992theory, eguchi1980gravitation}. Physically, quadrupoles of disclinations can be realized in liquid crystals, graphene sheets, and other condensed matter systems where topological defects influence mechanical and electronic properties. The study of quadrupoles is essential for understanding the response of materials to complex defect arrangements and their role in the modification of electronic wave functions, phonon spectra, and transport phenomena. Recent studies in spiral spin liquids have identified momentum vortices, interpreted as quadrupoles of disclinations in elasticity theory, as the main excitations governing the low-energy behavior of such systems~\cite{yan2021spiral}. These findings further reinforce the relevance of geometrically regularized quadrupolar configurations as fundamental topological excitations across various condensed matter contexts. The use of disclination quadrupoles as fundamental elements in modeling defects in continuous media is supported by both physical evidence and geometric reasoning. Suhanov et al.~\cite{suhanov2016} showed that, in metallic nanocrystals, quadrupolar configurations of partial disclinations offer greater energetic stability, more localized and isotropic stress fields, and improved elastic regularity compared to dipolar configurations.

In this work, we investigate the geometric and topological properties of quadrupolar configurations of disclinations and their impact on the self-energy and self-force of electric and magnetic sources. Quadrupolar configurations introduce richer geometric structures, significantly modifying the curvature fields and leading to distinct self-energy contributions depending on their spatial arrangement. We consider two representative cases: a linear quadrupole of disclinations, characterized by anisotropic curvature with a dominant dependence $\cos(2\theta)$ in the far-field regime; and a square quadrupole, generating a more isotropic curvature pattern that includes higher-order angular harmonics like $\cos(4\theta)$. By solving the Poisson equation for the conformal factor, we derive the induced curvature and Green's functions necessary for evaluating self-energy and self-force effects. Additionally, we examine the electric field structure and quantify its topological properties through numerical computation of winding numbers. Our results establish a clear relationship between the geometry introduced by quadrupolar defects, curvature-induced electromagnetic fields, and their associated topological characteristics. Finally, we discuss potential applications of our findings to condensed matter systems, including graphene, nematic materials, and gravitational analog models, emphasizing the broad interdisciplinary significance of structured topological defects.

This paper is organized as follows: In Section II, we present the conformal metric formulation for quadrupolar disclination configurations and derive the Green’s functions necessary to evaluate self-energy and self-force in linear sources. Section III details the calculation of electrostatic and magnetostatic self-energies and discusses their dependence on the symmetry and curvature induced by different quadrupole arrangements. In Section IV, we analyze the electric field structure and its topological properties, emphasizing the winding number associated with each quadrupole geometry. Finally, Section V summarizes our conclusions and highlights potential applications of these findings in condensed matter systems and gravitational analogs.

\section{Two-Dimensional Defects and Conformal Metric Approach}

Defects in lower-dimensional systems play a fundamental role in condensed matter physics, particularly in the geometric description of elastic media. Theoretical models establish an analogy between defect structures and gravitational fields, where disclinations and dislocations correspond to curvature and torsion, respectively~\cite{katanaev1992theory, kleinert1989gauge}. This analogy is particularly fruitful in (2+1)-dimensional systems, where gravity becomes a topological theory and allows for exact solutions that model defects via conical singularities~\cite{DeserJackiw1984, DeserJackiw1989}. In two-dimensional systems, the metric can always be written in a conformal form:
\begin{equation}
    ds_{2D}^{2}=\exp (2\Omega )\left( dr^{2}+r^{2}d\theta ^{2}\right),
    \label{mconforme}
\end{equation}
where $\Omega(r,\theta)$ is the conformal factor encoding the geometric distortions induced by defects. This formulation leads naturally to a Poisson equation for $\Omega$, and then we have that
\begin{equation}
    \Delta \Omega = -\lambda, 
    \label{poisson}
\end{equation}
we have that $\lambda$ represents the defect density, directly linked to the stress-energy tensor $T_{\alpha\beta}$. A complete geometric formulation that includes connections, curvature, and torsion can be developed using the formalism of Cartan geometry and vielbeins~\cite{eguchi1980gravitation}. The Ricci scalar for this metric is given bellow,
\begin{equation}
    R = 2e^{2\Omega} \Delta \Omega.
    \label{escalar-curv}
\end{equation}
Thus, the problem of describing defects reduces to solving the Poisson equation for $\Omega$, with different distributions of $\lambda$ corresponding to different defect configurations. This framework provides a powerful tool for modeling the geometry of defects in condensed matter systems, allowing for direct computation of induced curvature and associated physical effects. It also creates a bridge between elasticity theory and gravity in $(2+1)$-dimensions, making it possible to reinterpret stress and strain fields as manifestations of curvature and torsion. In the following sections, we analyze the quadrupole arrangements of disclinations and their impact on the local metric structure.

\section{Quadrupole of Disclinations}

The geometric theory of defects developed by Katanaev and Volovich provides a powerful framework for modeling disclinations and dislocations such as curvature and torsion singularities in effective Riemann-Cartan geometry~\cite{katanaev1992theory}. In this context, a disclination is modeled as a conical singularity, characterized by a deficit angle that directly corresponds to a localized Gaussian curvature. When two disclinations of opposite sign are brought together, they form a curvature dipole as a geometric analog of an edge dislocation. This configuration gives rise to a translational holonomy known as the Burgers vector, and can be interpreted as a nontrivial solution of the Einstein–Cartan equations in $(2+1)$ dimensions with torsion concentrated along a line. The metric induced by a disclination dipole reflects an anisotropic curvature field and serves as a prototype for more complex defect structures, such as quadrupoles. This approach not only unifies elasticity and differential geometry, but also allows for exact solutions in terms of conformal factors, making it well suited to the analysis of self-energy and self-force effects in media with distributed defects~\cite{katanaev1992theory, kleinert1989gauge}. Disclinations are sources of curvature and can be modeled using conical metrics, representing angular deficits or excesses in the material. Dislocations, in turn, emerge as the limiting case of a disclination dipole, where the net curvature cancels but torsion remains, encoding translational holonomy via the Burgers vector. Quadrupolar configurations of disclinations may lead to partial compensation of torsion or to the emergence of dislocation networks, depending on their symmetry and orientation. This provides a unified geometric framework that systematically relates curvature and torsion to different classes of topological defects, reinforcing the connection between elasticity and differential geometry.

A quadrupole of disclinations consists of four disclinations arranged symmetrically, such that their combined effect creates a more complex curvature pattern compared to a single disclination or a dipole of disclinations. This configuration can be used to model certain crystalline defects, as well as gravitational analogues in lower-dimensional gravity models. Quadrupoles arise naturally in the study of elastic deformations in solids and can be linked to the presence of stress fields with higher-order multipole structures. Unlike a single disclination, which introduces a local curvature singularity, or a dipole of disclinations, which creates a linear displacement field, a quadrupole configuration leads to more intricate deformations of the surrounding space. The quadrupole structure can be thought of as two symmetrically positioned dipoles, leading to a distortion field that varies with both radial distance and angular position. This formulation has been extensively explored in Katanaev's work~\cite{katanaev1992theory}, where the quadrupole configuration emerges as a fundamental structure in the geometric theory of defects.

In order to describe a quadrupolar configuration of disclinations, two representative geometries are considered: in the linear quadrupole, the disclinations are aligned along a straight line, generating an anisotropic stress field; in the square quadrupole, they are placed at the vertices of a square, leading to a more isotropic curvature distribution. Both arrangements are described by the Poisson equation for the conformal factor $\Omega(r, \theta)$, as previously introduced in Eq.~\eqref{poisson}, where $\lambda$ is the defect density. In the linear case, disclinations are symmetrically positioned along the $x$-axis at $(\pm d, 0)$ and $(\pm D, 0)$ with alternating topological charges $m_i$. The resulting conformal factor, obtained by solving the Poisson equation using the Green’s function of the Laplacian, takes the form
\begin{equation}
    \Omega(r,\theta) = - \sum_{i=1}^{4} \frac{m_i}{2\pi} \ln |\mathbf{r} - \mathbf{r}_i| + C.
\end{equation}
In the far-field limit, an angular expansion reveals a dominant quadrupolar structure:
\begin{equation}
    \Omega(r,\theta) \approx - \frac{Q}{r^2} \cos(2\theta),
\end{equation}
where $Q = \frac{4 m d^2}{\pi}$ is the effective quadrupole strength. The induced metric becomes
\begin{equation}
    ds^2 = e^{-2 Q \cos(2\theta)/r^2} \left( dr^2 + r^2 d\theta^2 \right),
\end{equation}
which can be approximated to first order for small deformations as
\begin{equation}
    ds^2 \approx \left(1 - \frac{2 Q}{r^2} \cos(2\theta)\right) \left( dr^2 + r^2 d\theta^2 \right).
\end{equation}
This configuration models a system with strong anisotropic curvature, where the geometric distortion is concentrated along the axis of the Burgers vectors. The linear quadrupole can also be interpreted as the superposition of two dislocations, each corresponding to a dipole of disclinations. In this case, the resulting configuration possesses a net Burgers vector given by the vectorial sum of the two constituent dislocations:
\begin{equation}
    \mathbf{b}_{\text{quad,lin}} = \mathbf{b}_1 + \mathbf{b}_2.
\end{equation}
The presence of a nonzero total Burgers vector implies residual torsion in the effective geometry, in addition to the curvature already discussed. This torsional contribution manifests as a net translational holonomy for closed loops encircling the configuration, distinguishing the linear quadrupole from purely curvature-inducing defects such as isolated disclinations. 

On the other hand, the square quadrupole consists of four disclinations located at $(\pm d, \pm d)$ with alternating charges such that $m_1 = -m_3$ and $m_2 = -m_4$. This choice ensures the cancellation of dipolar components and results in a more symmetric field. The defect density can be expanded as a multipolar series:
\begin{equation}
    \lambda(x,y) \approx \frac{Q}{(x^2 + y^2)^{5/2}} \left( a_0 + a_2 \cos(2\theta) + a_4 \cos(4\theta) \right),
\end{equation}
capturing the angular structure through coefficients $a_0$, $a_2$, and $a_4$. The Burgers vectors are arranged so that their resultant vector is null, minimizing net displacement and promoting an isotropic curvature field. Therefore, in contrast to the linear configuration, the square quadrupole has its Burgers vectors arranged such that they cancel pairwise as follow,
\begin{equation}
 \sum_{i=1}^{4} \mathbf{b}_i = 0.
\end{equation}
As a result, the global torsion vanishes, and the geometry does not exhibit a net translational holonomy. Although local curvature remains due to the disclinations, the cancellation of lower-order multipole moments ensures that the far-field geometry is dominated by higher angular harmonics, leading to a smoother and more isotropic curvature profile. This makes the square quadrupole a compelling example of a geometrically regularized defect with vanishing torsion and finite curvature. 

Solving the Poisson equation via the Green's function method in Cartesian coordinates for the square quadrupole configuration leads to the conformal factor, we have that
\begin{equation}
    \Omega(x,y) = \sum_{i=1}^{4} \frac{m_i}{2\pi} \ln \left( \sqrt{(x - x_i)^2 + (y - y_i)^2} \right).
\end{equation}
For $r \gg d$, an expansion gives us the follow expression
\begin{equation}
    \Omega(r,\theta) \approx - \frac{Q}{r^3} \left( a_0 + a_2 \cos(2\theta) + a_4 \cos(4\theta) \right),
\end{equation}
and the corresponding metric becomes
\begin{equation}
    ds^2 = e^{-2 Q \left(a_0 + a_2 \cos(2\theta) + a_4 \cos(4\theta)\right)/r^3} \left( dr^2 + r^2 d\theta^2 \right),
\end{equation}
which, for small perturbations, yields:
\begin{equation}
    ds^2 \approx \left(1 - \frac{2 Q}{r^3} \left(a_0 + a_2 \cos(2\theta) + a_4 \cos(4\theta)\right) \right) \left( dr^2 + r^2 d\theta^2 \right).
\end{equation}
The square configuration thus produces a smoother, more isotropic curvature profile, especially in the diagonal directions. This is a direct consequence of the symmetric arrangement of disclinations and the cancellation of lower-order multipole terms. These quadrupolar geometries offer a rich setting to study geometric self-interactions, as they interpolate between highly directional and more isotropic curvature fields, and can be understood as regularized versions of classical defects. This analysis is deeply rooted in the geometric theory of defects, where the curvature and torsion associated with disclinations and dislocations are captured through an effective Riemann–Cartan geometry~\cite{katanaev1992theory, kleinert1989gauge}. The use of conformal metrics in $(2+1)$-dimensions provides not only analytical tractability but also a natural framework for studying multipolar generalizations of line defects in a geometric language.

\section{Electrostatic and Magnetostatic Self-Energies in the Quadrupolar Background}

The study of the interaction between charged particles and topological defects has revealed subtle but significant effects on the electrostatic properties of systems in nontrivial media. In particular, the presence of defects such as disclinations or dislocations modifies the geometry of the space surrounding the charge, inducing electrostatic self-energy even in the absence of external sources. This phenomenon, known as geometry-induced self-interaction, arises from the boundary conditions imposed by the defects on classical fields and can be rigorously treated through the use of a regularized Green function associated with the underlying geometry. In this context, we consider the classical problem of a point charge at rest in a two-dimensional medium endowed with a conformal metric, associated with a continuous distribution of disclinations. By solving the Poisson equation in this geometric background, we obtain the self-energy of the charge as a direct consequence of the effective curvature of space, highlighting the deep connection between topology, geometry, and electrodynamics.

The expressions for electrostatic and magnetostatic self-energies follow classical formulations based on Green function techniques, as found in standard references~\cite{jackson1999classical, landau1975classical, panofsky1962classical, schwinger1998classical}. In the electrostatic case, we consider a linear distribution of the charge along an infinite wire parallel to the defect structure. Due to translational invariance, the Poisson equation reduces to two dimensions, and the scalar potential is given by,
\begin{equation}
\Phi (\vec{x}) = \int d^2x^{\prime} \rho (\vec{x}) G_b^{(2)}(\vec{x},\vec{x}^{\prime}),
\end{equation}
where \( G_b^{(2)}(\vec{x}, \vec{x}^{\prime}) \) is the two-dimensional Green function that solves the Poisson equation in the background geometry defined by the disclination quadrupole. This geometry is encoded in a conformal metric of the form \( g_{ab} = e^{-2\Omega(\vec{x})} \delta_{ab} \), where the function \( \Omega(\vec{x}) \) reflects the effect of the defect distribution. The Green function \( G_b^{(2)} \) satisfies
\begin{equation}
\Delta_b G_b^{(2)}(\vec{x}, \vec{x}^{\prime}) = \delta^{(2)}(\vec{x} - \vec{x}^{\prime}),
\end{equation}
and its regularized coincident limit contains all the geometric contributions to the self-energy. Assuming the charge distribution is modeled as a linear source, we obtain
\begin{equation}
\rho (\vec{x}) = \lambda \delta^{(2)} (\vec{x}-\vec{x}^{\prime}),
\label{charge}
\end{equation}
with linear charge density \( \lambda \), and then the electrostatic self-energy per unit length becomes:
\begin{equation}
\frac{U_{ele}}{l} = \frac{\lambda^2}{2} G_b^{(2)}(\vec{x},\vec{x}) \Big|_{reg},
\label{self_electrical}
\end{equation}
where \( G_b^{(2)}(\vec{x},\vec{x}) \big|_{reg} \) is the regularized Green function evaluated at coincident points, subtracting the usual flat-space divergence. This expression shows that the background geometry, encoded in the conformal factor and consequently in the Green function, modifies the self-energy even in the absence of external interactions. Such geometry-induced effects have analogues in quantum field theory, where background curvature leads to vacuum polarization phenomena~\cite{fumeronKR2022}.

The magnetostatic case is analogous, with the energy of a stationary current distribution presented in the follow way
\begin{equation}
U_{mag} = \frac{1}{2c} \int d^3x \, \vec{J}(\vec{x}) \cdot \vec{A}(\vec{x}),
\end{equation}
where \( \vec{J}(\vec{x}) \) is the current density and \( \vec{A}(\vec{x}) \) is the vector potential in the Coulomb gauge, satisfying \( \nabla \cdot \vec{A} = 0 \). Assuming a steady linear current along the \( z \)-direction as
\begin{equation}
\vec{J}(\vec{x}) = j_0 \hat{z} \delta^{(2)} (\vec{x}-\vec{x}^{\prime}),
\label{current}
\end{equation}
the magnetostatic self-energy per unit length can be written as
\begin{equation}
\frac{U_{mag}}{l} = \frac{j_0^2}{2c^2} G_b^{(2)}(\vec{x},\vec{x}) \Big|_{reg}.
\label{self_magnetic}
\end{equation}
The structural similarity between Eqs.~\eqref{self_electrical} and~\eqref{self_magnetic} emphasizes how both electric and magnetic self-interactions are governed by the same Green function, highlighting the geometric origin of these effects. This correspondence echoes results found in curved-space analogues of classical field theories, including Kalb-Ramond models~\cite{fumeronKR2022}.

\subsection{Green Function in the Quadrupole Metric}

Finding Green's functions in a non-Euclidean metric is, in general, a complex task. However, the quadrupole of disclinations exhibits an important property: it is conformally related to the Euclidean metric, meaning the spatial line element can be written in a simple way
\begin{equation}
g_{ab} = e^{-\Omega} \delta_{ab},  \label{metric}
\end{equation}
where $\Omega$ is the conformal factor associated with the defect configuration. Since the Green's functions satisfies the two-dimensional Poisson equation,
\begin{equation}
\Delta_E G(\vec{x}^{\prime},\vec{x}) = \delta^2(\vec{x}^{\prime} - \vec{x}),
\end{equation}
its solution in flat space is presented in the way
\begin{equation}
G_0^{(2)}(\vec{x}^{\prime},\vec{x}) = \frac{1}{4\pi} \ln |\vec{x}^{\prime} - \vec{x}|^2.
\end{equation}
Note that, in a conformally flat geometry, the coincident limit of the regularized Green function becomes~\cite{carvalho2000self, carvalho2004self}:
\begin{equation}
G_b^{(2)}(\vec{r}_0,\vec{r}_0) \Big|_{reg} = \Omega(\vec{x}).
\end{equation}
This important result has been rigorously demonstrated by Grats and Garcia~\cite{grats1998green}, who showed that for a conformally flat two-dimensional space, the regularized Green function at coincident points is proportional to the conformal factor. This proportionality significantly simplifies the computation.

\subsection{Self-Energy  in the Quadrupole Metric}

\begin{figure}[t]
    \centering
    \begin{subfigure}[b]{0.49\textwidth}
        \centering
        \includegraphics[width=\textwidth]{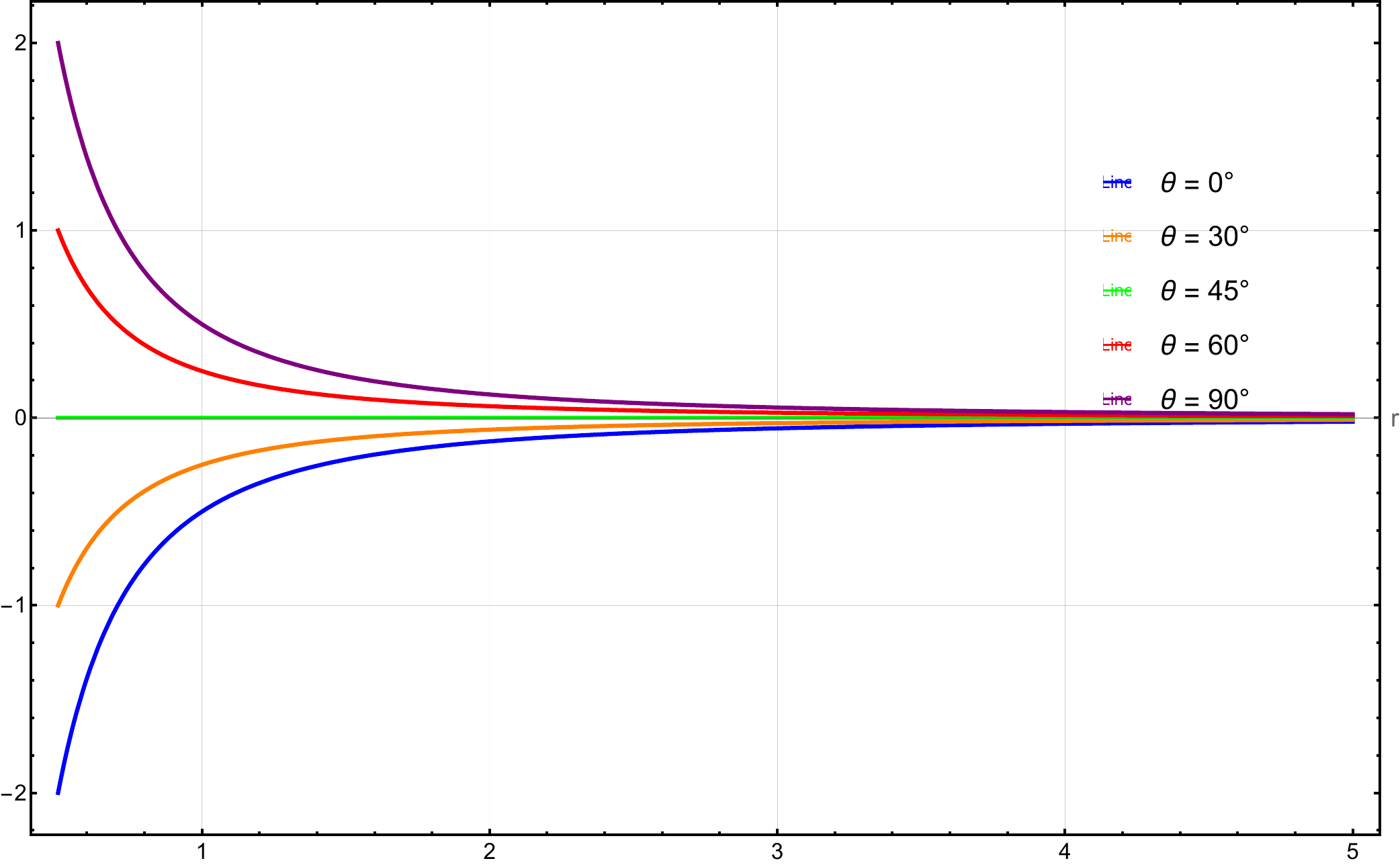}
        \caption{The angular dependence significantly influences the shape of the potential, with a maximum at \( \theta = 0^\circ \) and vanishing at \( \theta = 90^\circ \) for the linear quadrupole configuration.}
        \label{Ulinear}
    \end{subfigure}
    \hfill
    \begin{subfigure}[b]{0.49\textwidth}
        \centering
        \includegraphics[width=\textwidth]{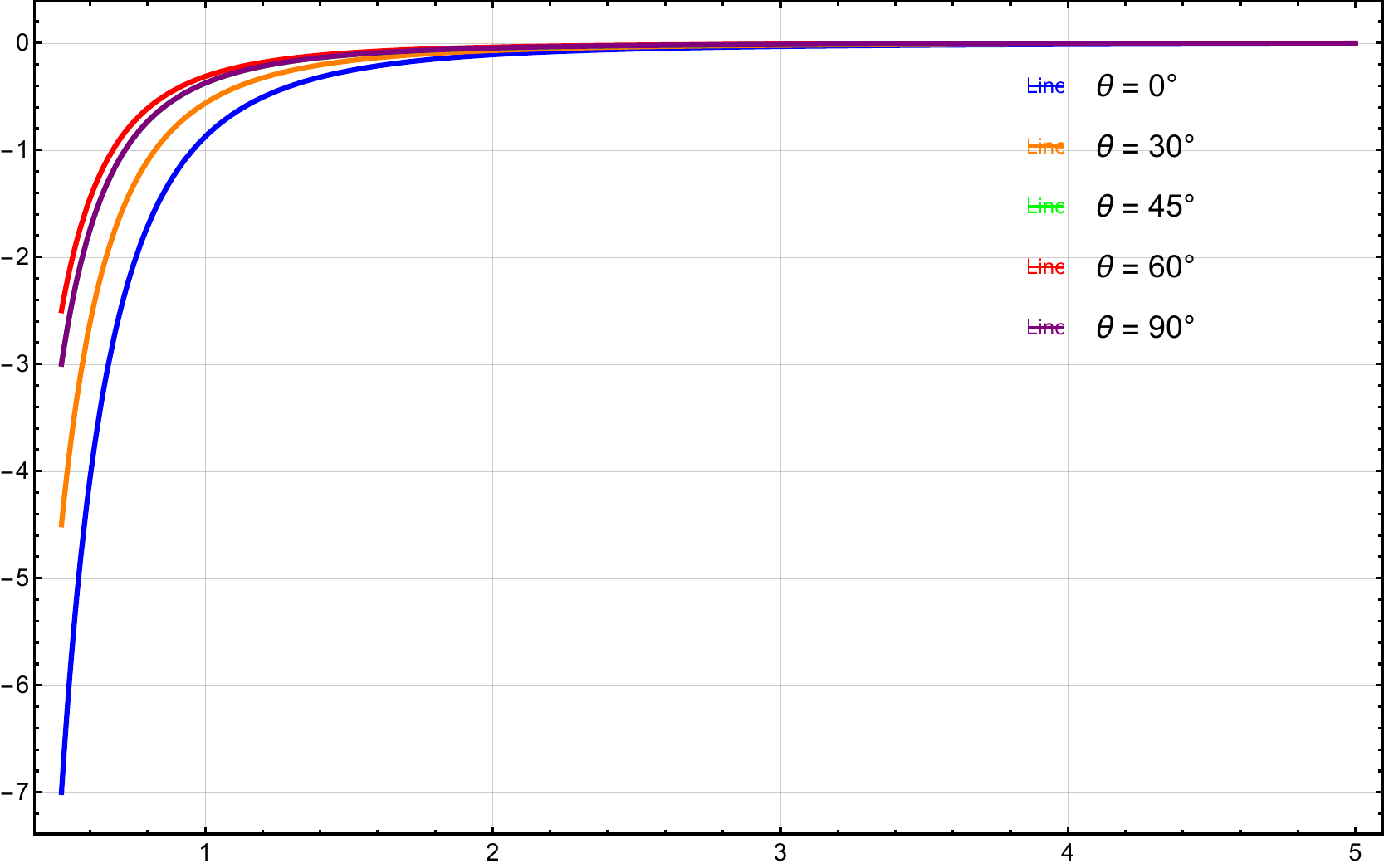}
        \caption{The self-energy exhibits negative values and depends significantly on the angular orientation, with a decay proportional to \( r^{-3} \), for the quadrupole square configuration.}
       \label{Uquadrado}
    \end{subfigure}
  \caption{Plot of the self-energy \( U(r) \) of each quadrupole configuration as a function of the radial distance \( r \), for different angles \( \theta \in \{0^\circ, 30^\circ, 45^\circ, 60^\circ, 90^\circ\} \).}
\end{figure}

For the quadrupole metric, the conformal factor $\Omega$ depends on the configuration of the defect. In the linear quadrupole arrangement, as we previously showed, the conformal factor can be write by
\begin{equation}
\Omega_L = -\frac{Q}{r^2} \cos(2\theta),
\end{equation}
while in the square quadrupole configuration, it becomes
\begin{equation}
\Omega_S = -\frac{Q}{r^3} \left( a_0 + a_2 \cos(2\theta) + a_4 \cos(4\theta) \right).
\end{equation}
Substituting into the self-energy expressions yields:
\begin{subequations}
\begin{eqnarray}
\frac{U_{ele}}{l} &=& \frac{\lambda^2}{2} \Omega(\vec{x}), \\
\frac{U_{mag}}{l} &=& \frac{j_0^2}{2c^2} \Omega(\vec{x}).  
\end{eqnarray}
\end{subequations}

To illustrate the anisotropy of the electrostatic self-energy induced by the geometry of a linear quadrupole of disclinations, we construct angular profiles based on the analytical form of the self-energy. These profiles, plotted in polar coordinates, represent the variation of self-energy as a function of angle \( \theta \), for a fixed radius \( r \). The expression used explicitly incorporates the physical constants of the system, such as the charge \( q \), the linear charge density \( \lambda \), and the angular factor \( d \), which encodes the symmetry of the defect. The radial deformation of the circle reveals directions of higher and lower self-energy, highlighting the role of geometric curvature in the self-interaction of the charge. This profile can be interpreted as an \textit{angular scalar field} induced by the geometry, which describes the intensity of self-energy in each direction. In what follows, we present the corresponding plots for different symmetries.

To better understand the influence of defect symmetry on self-interactions, we analyze the electrostatic self-energy profiles associated with linear and square quadrupole configurations of disclinations. Figures~\ref{Ulinear} and~\ref{Uquadrado} present the regularized self-energy \( U(r) \) as a function of the radial coordinate \( r \), for several angular directions \( \theta \in \{0^\circ, 30^\circ, 45^\circ, 60^\circ, 90^\circ\} \). The linear configuration (Fig.~\ref{Ulinear}) exhibits a strongly anisotropic potential profile dominated by an angular dependence of \( \cos(2\theta) \), with energy maxima and minima aligned along the principal axes. This results in a directional modulation of the curvature field and a radial decay of the form \( r^{-2} \), characterizing long-range interactions. In contrast, the square configuration (Fig.~\ref{Uquadrado}) incorporates higher-order angular harmonics such as \( \cos(4\theta) \), leading to a more isotropic four-lobe structure. The radial decay in this case follows \( r^{-3} \), indicating a more localized interaction near the defect core. Furthermore, the square quadrupole presents a vanishing net Burgers vector, which suppresses torsion and enhances geometric regularity. This comparison highlights the role of symmetry and multipolar structure in shaping the angular and radial behavior of curvature-induced self-interactions in two-dimensional defect systems.

\begin{figure}[t]
    \centering
    \begin{subfigure}[b]{0.49\textwidth}
        \centering
        \includegraphics[width=\textwidth]{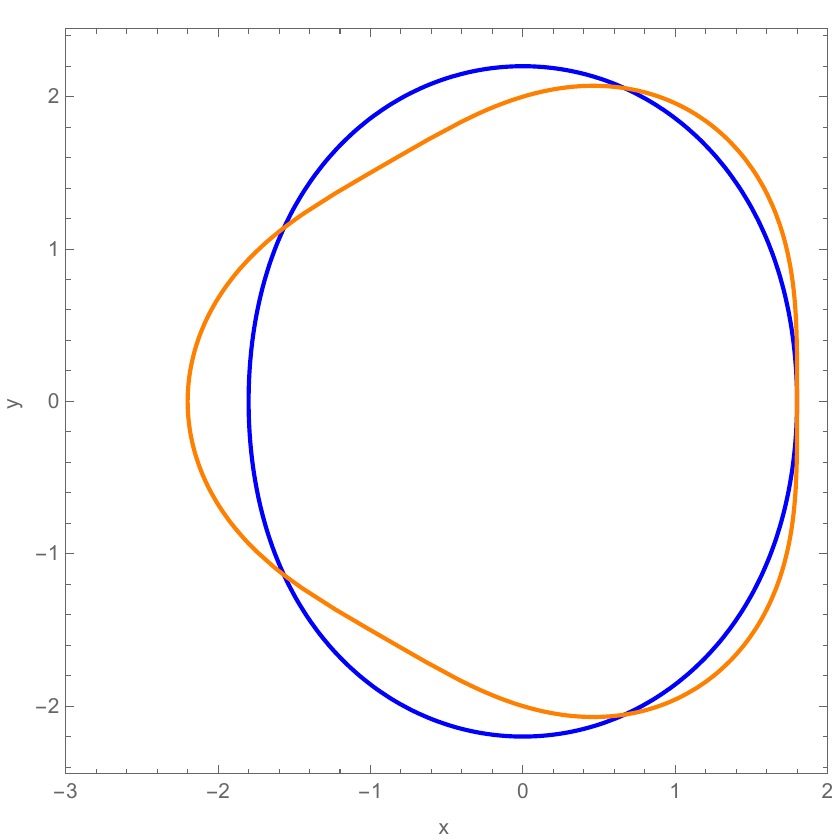}
        \caption{The plot shows the angular dependence of the self-energy at fixed radius, revealing alternating regions of attraction and repulsion modulated by the defect geometry. A characteristic axial symmetry emerges from the linear arrangement of disclinations.}
        \label{perfil.linear}
    \end{subfigure}
    \hfill
    \begin{subfigure}[b]{0.449\textwidth}
        \centering
        \includegraphics[width=\textwidth]{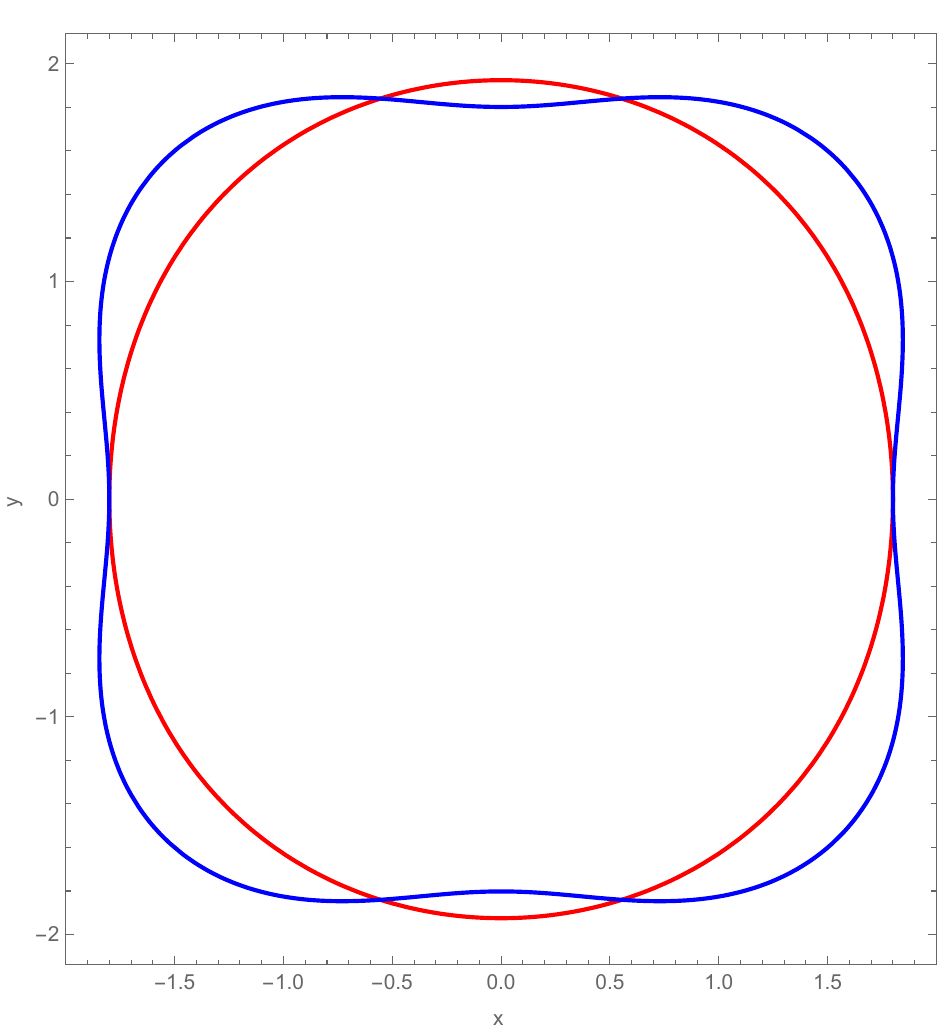}
        \caption{The plot displays two curves at fixed radius and \( a_2 = 0.5 \), corresponding to distinct values of \( a_4 \): the red curve represents \( a_4 = -0.1 \), while the blue curve corresponds to \( a_4 = -1 \). The shapes of the curves reflect how higher-order angular harmonics modulate the anisotropy of the self-energy.}
       \label{perfil.quadrado}
    \end{subfigure}
  \caption{Angular profile of the regularized electrostatic self-energy for a test charge in the presence of each quadrupole configuration of disclinations.}
\end{figure}

To further illustrate the anisotropic nature of the electrostatic self-energy induced by quadrupolar disclination configurations, we construct angular profiles based on the analytical expressions of the regularized self-energy. These profiles are plotted in polar coordinates and represent the variation in self-energy as a function of the angular coordinate \( \theta \), at a fixed radial distance \( r \). The expressions retain explicit physical parameters, including the test charge \( q \), the linear charge density \( \lambda \), and the angular coefficients \( a_2 \) and \( a_4 \), which encode the symmetry properties of each defect configuration. These profiles can be interpreted as angular scalar fields defined on a circle, describing the directional modulation of the geometry-induced self-interaction. Deviations from the perfect circular shape directly reflect the angular structure of the underlying curvature. In what follows, we present representative plots for both the linear and square quadrupole configurations.

The figure~\ref{perfil.linear} shows a well-defined axial symmetry, consistent with the linear arrangement of disclinations along a principal axis. The resulting two-lobe structure originates from the dominant term \( \cos(2\theta) \) in the conformal factor. This angular dependence generates alternating directions of enhanced and suppressed self-energy, revealing anisotropic features in the effective interaction landscape. This behavior underscores the role of geometry in shaping the directional characteristics of electrostatic self-interactions. 

On the other hand, in figure~\ref{perfil.quadrado}, the inclusion of the higher-order angular harmonic \( \cos(4\theta) \) enriches the angular modulation of the self-energy profile. The red and blue curves illustrate the effect of varying the parameter \( a_4 \), while keeping \( a_2 \) fixed. The blue curve, associated with the most negative value \( a_4 = -1 \), exhibits sharper angular features and increased anisotropy, reflecting a stronger geometric modulation. In contrast, the red curve (\( a_4 = -0.1 \)) presents a smoother and more isotropic profile. This comparison highlights how the square quadrupole configuration enables fine-tuning of curvature and angular anisotropy through geometric parameters, offering a versatile framework for modeling directional self-interactions in media with topological defects.

\subsection{Self-forces in Conformal Geometries}

Self-forces emerge naturally from the position dependence of the regularized self-energies. Since the background metric is conformally flat, the gradient must be interpreted with respect to the curved geometry. For a linear charge distribution with density \( \lambda \), the electrostatic self-force per unit length reads
\begin{equation}
\frac{\vec{F}_{\text{ele}}}{l} = -\nabla_g \left( \frac{U_{\text{ele}}}{l} \right) = -\frac{\lambda^2}{2} \nabla_g \Omega(\vec{x}),
\end{equation}
where \( \nabla_g \) denotes the gradient compatible with the conformal metric. In two-dimensional conformal geometries, it can be related to the Euclidean gradient as
\begin{equation}
\nabla_g = e^{-\Omega(\vec{x})} \nabla_E,
\end{equation}
which leads to the general expressions for the self-forces per unit length:
\begin{subequations}
\begin{align}
\frac{\vec{F}_{\text{ele}}}{l} &= -\frac{\lambda^2}{2} \, e^{-\Omega(\vec{x})} \nabla_E \Omega(\vec{x}), \\
\frac{\vec{F}_{\text{mag}}}{l} &= +\frac{j_0^2}{2c^2} \, e^{-\Omega(\vec{x})} \nabla_E \Omega(\vec{x}).
\end{align}
\end{subequations}
These expressions highlight that both self-forces are entirely determined by the spatial structure of the conformal factor \( \Omega(\vec{x}) \), which encodes the geometric deformation induced by the underlying topological defect. The direction and intensity of the resulting forces reflect the local curvature gradients, indicating effective attraction or repulsion depending on the configuration.

We now compute the explicit self-forces for the two cases of interest: linear quadrupole and square quadrupole. For the linear quadrupole
\begin{subequations}
\begin{align}
\frac{\vec{F}_{\text{ele}}^{\text{lin}}}{l} &= -\frac{\lambda^2 Q}{r^3} \, e^{\frac{Q}{r^2} \cos(2\theta)} \left[ \hat{r} \cos(2\theta) + \hat{\theta} \sin(2\theta) \right], \\
\frac{\vec{F}_{\text{mag}}^{\text{lin}}}{l} &= +\frac{j_0^2 Q}{c^2 r^3} \, e^{\frac{Q}{r^2} \cos(2\theta)} \left[ \hat{r} \cos(2\theta) + \hat{\theta} \sin(2\theta) \right].
\end{align}
\end{subequations}
And for a square qudrupole the self-forces are
\begin{subequations}
\begin{align}
\frac{\vec{F}_{\text{ele}}^{\text{quad}}}{l} &= -\frac{\lambda^2}{2} \, e^{\frac{Q}{r^3} (a_0 + a_2 \cos(2\theta) + a_4 \cos(4\theta))} \times \nonumber \\
& \quad \left[ \hat{r} \cdot \frac{3Q}{r^4}(a_0 + a_2 \cos(2\theta) + a_4 \cos(4\theta)) + \hat{\theta} \cdot \frac{Q}{r^4} (2a_2 \sin(2\theta) + 4a_4 \sin(4\theta)) \right], \\
\frac{\vec{F}_{\text{mag}}^{\text{quad}}}{l} &= +\frac{j_0^2}{2c^2} \, e^{\frac{Q}{r^3} (a_0 + a_2 \cos(2\theta) + a_4 \cos(4\theta))} \times \nonumber \\
& \quad \left[ \hat{r} \cdot \frac{3Q}{r^4}(a_0 + a_2 \cos(2\theta) + a_4 \cos(4\theta)) + \hat{\theta} \cdot \frac{Q}{r^4} (2a_2 \sin(2\theta) + 4a_4 \sin(4\theta)) \right].
\end{align}
\end{subequations}
These forces reflect the complex interplay between geometry and field self-interaction, with angular dependence dictated by the quadrupolar symmetry of the underlying distribution. In particular, the radial and angular components encode anisotropic effects that are highly sensitive to both distance and orientation relative to the defect's structure.
\begin{figure}[t]
    \centering
    \begin{subfigure}[b]{0.49\textwidth}
        \centering
        \includegraphics[width=\textwidth]{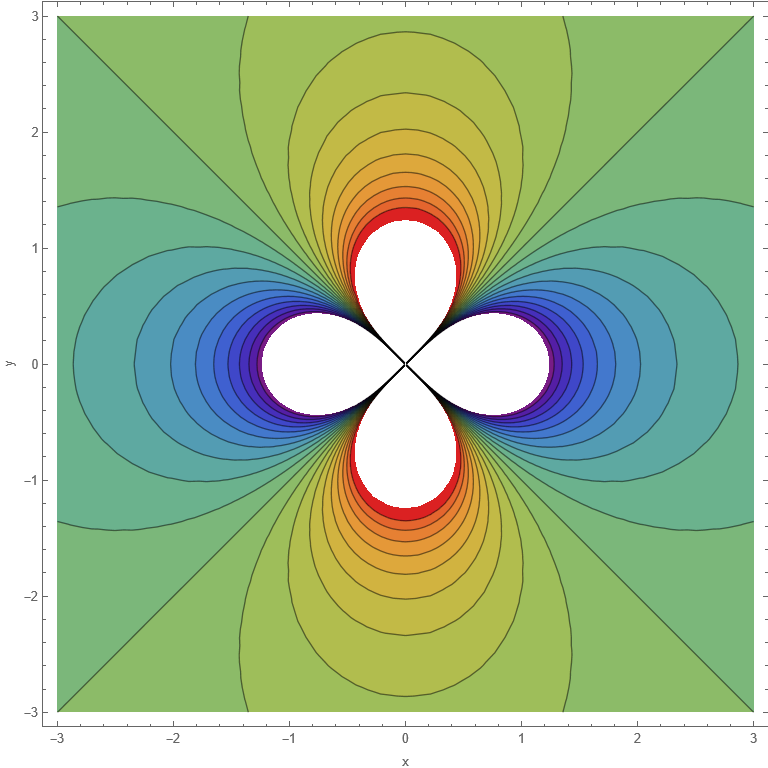}
        \caption{2D equipotential lines.}
    \end{subfigure}
    \hfill
    \begin{subfigure}[b]{0.49\textwidth}
        \centering
        \includegraphics[width=\textwidth]{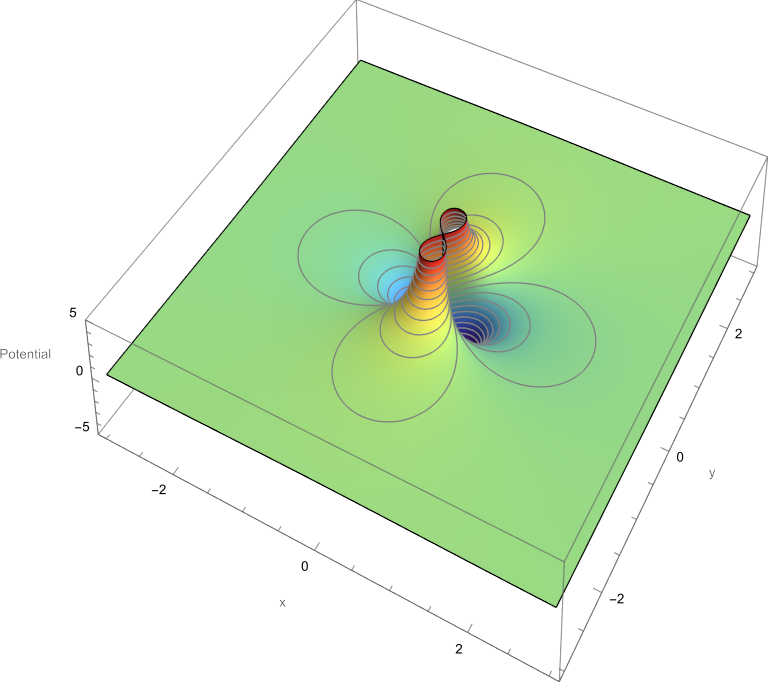}
        \caption{3D potential surface.}
    \end{subfigure}
  \caption{Equipotential surfaces for the linear quadrupole configuration. The 2D plot shows anisotropic lobes aligned with the principal axes due to the $\cos(2\theta)$ dependence. The 3D surface highlights the directional variation of the curvature-induced potential.}
    \label{fig:equipotencial-linear}
\end{figure}
The spatial dependence of the regularized self-energies naturally leads to the notion of equipotential surfaces, defined as the loci where the energy remains constant. Since the self-energies are directly proportional to the conformal factor, these surfaces correspond to level sets of the conformal function. In other words, the geometry of the defect induces a potential landscape whose structure can be fully visualized through the contours of constant energy. For the linear quadrupole, these equipotential surfaces exhibit a characteristic dipolar symmetry, with lobes aligned along diagonal axes. In contrast, the square quadrupole introduces richer angular patterns due to its four-fold symmetry, resulting in more intricate equipotential structures. These geometric features illustrate not only the curvature induced by the defect but also the anisotropy of the resulting self-interactions. The equipotential surfaces of the linear quadrupole, presented in Figure~\ref{fig:equipotencial-linear}, reveal a strongly anisotropic structure consistent with the leading-order angular dependence of the conformal factor, proportional to $\cos(2\theta)$. This angular modulation results in a two-lobed potential landscape, with maxima and minima aligned along the principal axes of the configuration. The pronounced contrast between regions of high and low curvature reflects the directional variation of the geometric deformation, with enhanced self-interaction along the horizontal and vertical directions, and suppressed interaction along the diagonals. This anisotropy is a direct manifestation of the quadrupolar curvature field, and encodes the symmetry and orientation of the underlying disclination arrangement, influencing both the magnitude and orientation of the resulting self-forces.

\begin{figure}[H]
    \centering
    \begin{subfigure}[b]{0.49\textwidth}
        \centering
        \includegraphics[width=\textwidth]{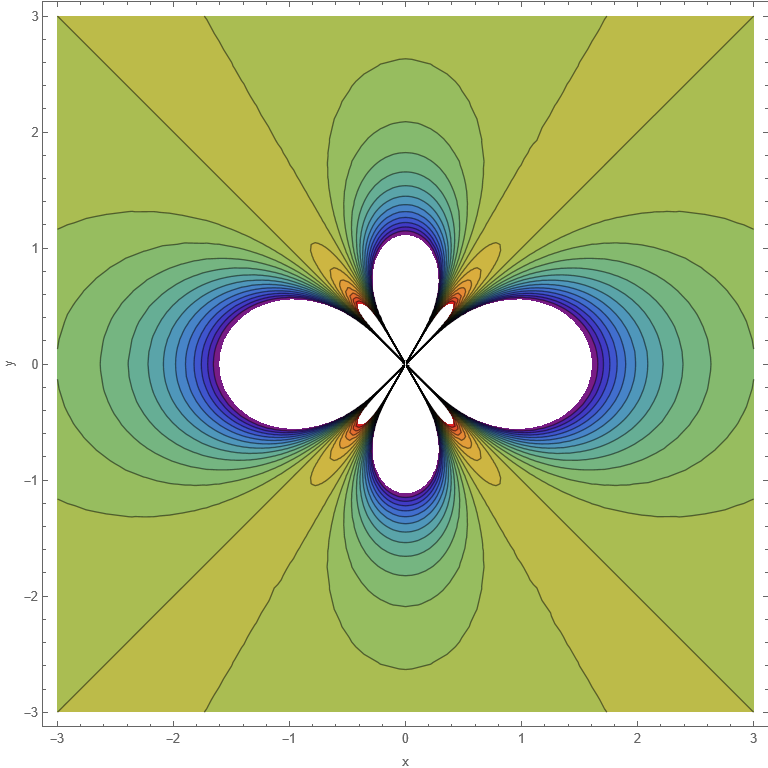}
        \caption{Square quadrupole — 2D equipotential lines.}
        \label{fig:equipotencial-quadrada-2d}
    \end{subfigure}
    \hfill
    \begin{subfigure}[b]{0.49\textwidth}
        \centering
        \includegraphics[width=\textwidth]{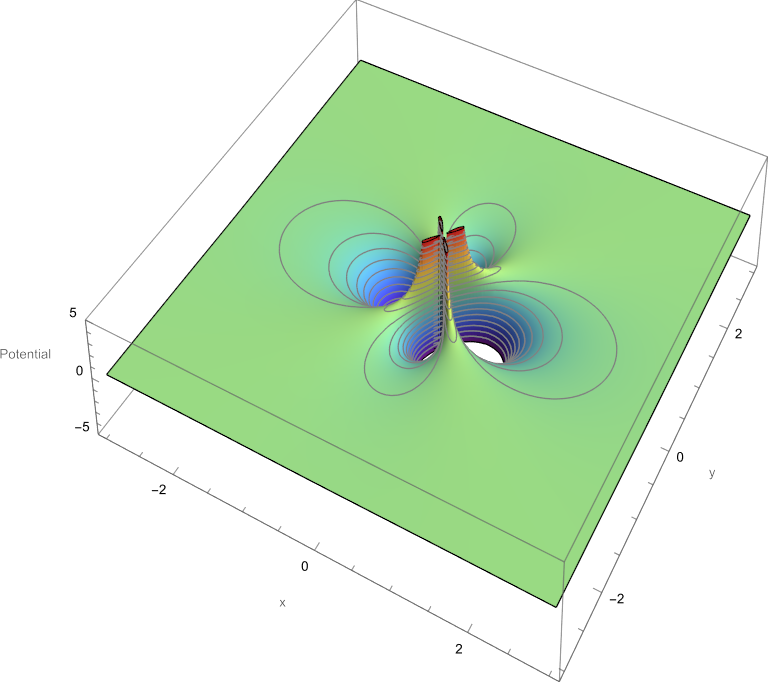}
        \caption{Square quadrupole — 3D potential surface.}
        \label{fig:equipotencial-quadrada-3d}
    \end{subfigure}
   \caption{Equipotential surfaces for the square quadrupole configuration. The 2D plot shows fourfold symmetry from higher-order harmonics, while the 3D surface reveals a smoother, more isotropic potential.}

    \label{fig:equipotencial-quadrada}
\end{figure}
 
The equipotential surfaces of the square quadrupole, shown in Figure~\ref{fig:equipotencial-quadrada}, display a higher degree of angular symmetry and isotropy, arising from the cancellation of lower-order multipole moments and the dominance of higher-order harmonics, particularly the $\cos(4\theta)$ term in the conformal factor. The resulting four-lobe structure reflects a potential landscape with smoother angular variation and a more homogeneous spatial distribution of curvature. Unlike the linear configuration, where the potential exhibits sharp directional gradients, the square quadrupole generates a field with reduced anisotropy, indicative of a balanced defect configuration with vanishing net Burgers vector. This leads to a more uniform self-interaction profile around the core defect, and suggests that curvature effects, while still present, are distributed in a more isotropic fashion, minimizing directional biases in the induced self-forces.

\section{Conclusion}

In this work, we investigated the geometric and physical consequences of quadrupolar disclination configurations in $(2+1)$-dimensional spacetime using a conformal metric framework. Two representative geometries were considered: a linear quadrupole, characterized by a dominant angular dependence $\cos(2\theta)$ and a square quadrupole, incorporating higher-order harmonics such as $\cos(4\theta)$, resulting in improved angular symmetry and isotropy. Analytical solutions for the conformal factor were obtained by solving the Poisson equation for each configuration. These expressions enabled the construction of Green functions and the computation of electrostatic and magnetostatic self-energies and self-forces associated with line sources.

The resulting self-forces reveal distinct physical behaviors: the linear quadrupole produces anisotropic forces with a slower radial decay ($r^{-3}$), while the square quadrupole yields more isotropic and short-range interactions ($r^{-4}$), consistent with the cancellation of lower-order multipole contributions. Moreover, the sign inversion between electric and magnetic self-forces reflects the geometric duality between charges and steady currents in curved backgrounds.

In addition to the analytic results, we presented visualizations of the angular profiles of the regularized self-energy for both linear and square quadrupoles (see Figures~\ref{perfil.linear} and \ref{perfil.quadrado}). These plots revealed how the angular symmetry of the defect structure governs the intensity of the self-interaction in different directions. The linear configuration enhances curvature anisotropy along the principal axes, while the square configuration smooths out curvature variations and introduces tunable anisotropy through the parameter \( a_4 \).

Altogether, our findings provide a coherent framework for understanding curvature-induced self-interactions and their dependence on geometric and angular symmetries. The approach developed here may be extended to higher-order multipolar distributions, time-dependent sources, or thermal effects. Future investigations may explore applications in condensed matter systems with effective geometric descriptions, including graphene, nematic liquid crystals, and topological phases of matter.

{\bf Acknowledgments:} This work was supported by Conselho Nacional de Desenvolvimento Cient\'{\i}fico e Tecnol\'{o}gico (CNPq) and Funda\c{c}\~ao de Apoio a Pesquisa do Estado da Para\'iba (Fapesq-PB). G. Q. Garcia would like to thank Fapesq-PB for financial support (Grant BLD-ADT-A2377/2024). The work by C. Furtado has been supported by the CNPq (project PQ Grant 1A No. 311781/2021-7).

\bibliographystyle{iopart-num}

  \bibliography{biblio}

\providecommand{\newblock}{}
\begin{thebibliography}{10}
\expandafter\ifx\csname url\endcsname\relax
  \def\url#1{{\tt #1}}\fi
\expandafter\ifx\csname urlprefix\endcsname\relax\def\urlprefix{URL }\fi
\providecommand{\eprint}[2][]{\url{#2}}

\bibitem{DeserJackiw1984}
Deser S and Jackiw R 1984 {\em Annals of Physics\/} {\bf 153} 405--416

\bibitem{DeserJackiw1989}
Deser S and Jackiw R 1989 {\em Annals of Physics\/} {\bf 192} 352--367

\bibitem{katanaev1992theory}
Katanaev M~O and Volovich I~V 1992 {\em Annals of Physics\/} {\bf 216} 1--28

\bibitem{Fumeron2021}
Fumeron S, Berche B and Moraes F 2021 {\em Liquid Crystals Reviews\/} {\bf 9} 85--110

\bibitem{kleinert1989gauge}
Kleinert H 1989 {\em Gauge Fields in Condensed Matter, Vol. II: Stresses and Defects\/} (World Scientific)

\bibitem{gopalakrishnan2006curvature}
Gopalakrishnan M, Golubović L and Lubensky T~C 2006 {\em Phys. Rev. E\/} {\bf 74} 010701(R)

\bibitem{lin}
Linet B 1986 {\em Phys. Rev. D\/} {\bf 33} 1833--1835

\bibitem{sou}
Souradeep T and Sahni V 1992 {\em Phys. Rev. D\/} {\bf 46} 1616--1631

\bibitem{mex}
Guimarães M~E~X and Linet B 1993 {\em Class. Quantum Grav.\/} {\bf 10} 1665--1668

\bibitem{eug1}
de~Mello E~R~B and Furtado C 1997 {\em Phys. Rev. D\/} {\bf 56} 1345--1348

\bibitem{mello2012}
de~Mello E~R~B 2012 {\em Phys. Rev. D\/} {\bf 85} 044020

\bibitem{furtado1997}
Furtado C and Moraes F 1997 {\em Class. Quantum Grav.\/} {\bf 14} 3425--3432

\bibitem{dirac1931quantised}
Dirac P~A~M 1931 {\em Proceedings of the Royal Society of London. Series A\/} {\bf 133} 60--72

\bibitem{wu1975concept}
Wu T~T and Yang C~N 1975 {\em Physical Review D\/} {\bf 12} 3845--3857

\bibitem{eguchi1980gravitation}
Eguchi T, Gilkey P~B and Hanson A~J 1980 {\em Physics Reports\/} {\bf 66} 213--393

\bibitem{dennis2009singular}
Dennis M~R, O'Holleran K and Padgett M~J 2009 {\em Progress in Optics\/} {\bf 53} 293--363

\bibitem{berry2009optical}
Berry M~V 2009 {\em Journal of Optics A: Pure and Applied Optics\/} {\bf 11} 094001

\bibitem{abulafia2023}
Abulafia Y, Goft A, Orion N and Akkermans E 2023 {\em Physical Review Letters\/} {\bf 130} 223801

\bibitem{volovik1998simulation}
Volovik G 1998 {\em Low Temperature Physics\/} {\bf 24} 127--129

\bibitem{carvalho2025scaterring}
Carvalho A~M~d~M, Marques A~S, Silva G~T, Garcia G~Q and Furtado C~Q 2025 {\em International Journal of Geometric Methods in Modern Physics\/}

\bibitem{candemir2023linear}
Candemir N and {\"O}zdemir A 2023 {\em Physics Letters A\/} {\bf 492} 129226

\bibitem{vozmediano2010gauge}
Vozmediano M~A, Katsnelson M and Guinea F 2010 {\em Physics Reports\/} {\bf 496} 109--148

\bibitem{mesaros2010parallel}
Mesaros A, Sadri D and Zaanen J 2010 {\em Physical Review B—Condensed Matter and Materials Physics\/} {\bf 82} 073405

\bibitem{mesaros2009berry}
Mesaros A, Sadri D and Zaanen J 2009 {\em Physical Review B—Condensed Matter and Materials Physics\/} {\bf 79} 155111

\bibitem{Guinea2010}
Guinea F, Katsnelson M~I and Geim A~K 2010 {\em Nature Physics\/} {\bf 6} 30--33

\bibitem{yan2021spiral}
Yan H and Reuther J 2021 {\em arXiv preprint arXiv:2112.10676\/} Also available as Phys. Rev. B 105, 224405 (2022) (\textit{Preprint} \eprint{2112.10676})

\bibitem{suhanov2016}
Suhanov I~I, Ditenberg I~A and Tyumentsev A~N 2016 Theoretical analysis of features of dipole and quadrupole configurations of partial disclinations in nanocrystals of metals {\em IOP Conference Series: Materials Science and Engineering\/} vol 116 (IOP Publishing) p 012035 \urlprefix\url{https://doi.org/10.1088/1757-899X/116/1/012035}

\bibitem{jackson1999classical}
Jackson J~D 1999 {\em Classical Electrodynamics\/} 3rd ed (Wiley)

\bibitem{landau1975classical}
Landau L~D and Lifshitz E~M 1975 {\em The Classical Theory of Fields\/} 4th ed ({\em Course of Theoretical Physics\/} vol~2) (Pergamon Press)

\bibitem{panofsky1962classical}
Panofsky W~K~H and Phillips M 1962 {\em Classical Electricity and Magnetism\/} 2nd ed (Addison-Wesley)

\bibitem{schwinger1998classical}
Schwinger J, DeRaad L~L, Milton K~A and Tsai W~Y 1998 {\em Classical Electrodynamics\/} (Perseus Books)

\bibitem{fumeronKR2022}
Berche B, Fumeron S and Moraes F 2022 {\em arXiv preprint\/} (\textit{Preprint} \eprint{2205.05295}) \urlprefix\url{https://arxiv.org/abs/2205.05295}

\bibitem{carvalho2000self}
de~M~Carvalho A~M, Furtado C and Moraes F 2000 {\em Phys. Rev. D\/} {\bf 62} 067504

\bibitem{carvalho2004self}
de~M~Carvalho A~M, Furtado C and Moraes F 2004 {\em Int. J. Mod. Phys. A\/} {\bf 19} 3039--3048

\bibitem{grats1998green}
Grats Y~V and Garcia A~A 1998 {\em Physical Review D\/} {\bf 58} 085007 \urlprefix\url{https://doi.org/10.1103/PhysRevD.58.085007}

\end{thebibliography}

  \end{document}